\documentstyle[12pt]{article}

\begin{document}

\begin{flushright}
gr-qc/0012051 \\
%$~$ \\
%14 December 2000\\
%(revised 26 December 2000)\\
December 2000\\
\end{flushright}

\begin{centering}
\bigskip
{\leftskip=2in \rightskip=2in
{\large \bf Relativity in space-times
with short-distance structure governed by an
observer-independent (Planckian) length scale}}\\
\bigskip
\bigskip
\bigskip
{\bf Giovanni AMELINO-CAMELIA}\\
\bigskip
Dipartimento di Fisica, Universit\'{a} ``La Sapienza", P.le Moro 2,
I-00185 Roma, Italy\\ 
\end{centering}

\vspace{1cm}
\begin{center}
{\bf ABSTRACT}
\end{center}

{\leftskip=0.6in \rightskip=0.6in

I show that it is possible to formulate the Relativity postulates 
in a way that does not lead to inconsistencies in the case of
space-times whose short-distance structure is 
governed by an observer-independent length scale.
The consistency of these
postulates proves incorrect the expectation that modifications of the
rules of kinematics involving the Planck length would necessarily
require the introduction of a preferred class of inertial observers.
In particular, it is possible for every inertial 
observer to agree on physical laws supporting deformed
dispersion relations of the type $E^2- c^2 p^2- c^4 m^2 + f(E,p,m;L_p)=0$,
at least for certain types of $f$.
}%%%%%

\newpage
%GAC%
\baselineskip 12pt plus .5pt minus .5pt
\pagenumbering{arabic}
\pagestyle{plain} 

\section{Introduction}
The (reduced) Planck constant $\hbar$ ($\hbar \simeq 10^{-34}J s$)
entered physics 100 years ago, when Planck announced
his solution of the black-body-radiation paradox.
As we look back~\cite{jackqm100,zeilqm100,gacqm100} 
at this century of physics with $\hbar$
we see a journey filled with the discovery of more and more
roles for this fundamental constant.
However, we have not yet established which role (if any)
should be played in the structure of space-time
by one of the implications of the
existence of $\hbar$ which appeared to be most significant to Planck:
the possibility to define the length scale now called 
Planck length $L_p$ by combining $\hbar$ with the 
gravitational constant $G$ and the speed-of-light constant $c$
($L_p \equiv \sqrt{\hbar G/c^3} \sim 1.6 {\cdot} 10^{-35}m$).
A significant obstruction for a physical interpretation of
the Planck length became manifest less than 5 years after Planck's
announcement, when Einstein introduced his powerful postulates
of Special Relativity, which in particular implied that
different values should be attributed to a physical length scale by
different (inertial) observers.
How could $L_p$ play a role in the structure of space-time
without violating Special Relativity?

This question has recently become more urgent, since certain classes
of experiments~\cite{ehns,grbgac,gacgwi,kifu,aus,ahlunature,polonpap} 
have reached sensitivity levels sufficient for testing phenomenological
models predicting new effects with strength set by the Planck length
(the magnitude of these effects is primarily governed by an overall
coefficient that can be described as some power of the ratio between the
Planck length and a length scale characteristic of the physical context).
An interesting example, which appears 
to be the phenomenological model that can be tested more 
accurately~\cite{grbgac,schaef,billetal,ita,gactp1}, is the one 
of deformations of the dispersion relation of the 
type $E^2- c^2 p^2- c^4 m^2+f(E,p,m;L_p)=0$. 
Some puzzling ``threshold anomalies"~\cite{gactp1} observed in two 
classes of astrophysics data~\cite{kifu,aus}
have motivated several authors (see, {\it e.g.}, 
Refs.~\cite{kifu,aus,ita,gactp1,colgla,kluz,sato})
to the study of deformations
of the dispersion relation.
Interestingly, under the assumption that the observed threshold anomalies
are due to a $E^2- c^2 p^2- c^4 m^2+f(E,p,m;L_p)=0$ deformation, when taking
into account the upper bounds of $f$ that have already been established
experimentally~\cite{schaef,billetal}, one is naturally led~\cite{gactp2}
to consider the possibility that the dimensionless quantity
setting the magnitude of $f$ is of order\footnote{Note that 
from this point onward I use
conventions with $\hbar = 1$.}
$L_p E/c$. 
This would also explain why there is no evidence of dispersion-relation 
deformation in low-energy data, where $L_p E/c$ has very small value, while
astrophysics data probing high-energy regimes could start to show traces
of the deformation. But this brings us back to the original question:
how could inertial observers agree on physical laws stating
that some new effects 
(say, those with $L_p E/c$ behaviour) 
can no longer be neglected
at a characteristic energy scale $c/L_p$ if a particle
with energy $c/L_p$ for one inertial observer can have 
much lower energy for another inertial observer?

Clearly in order to accommodate a fundamental space-time role for $L_p$
({\it e.g.} of a type leading to $E^2- c^2 p^2- c^4 m^2+f(E,p,m;L_p)=0$
with observer-independent value of $L_p$)
the Relativity relations between observations performed by
different inertial observers must also somehow encode some
information on $L_p$. This of course is not possible without
some modification of Einsten's Relativity postulates.
In this paper I show that it is possible to formulate rather natural
modifications of those postulates that could provide an
answer to the questions raised in this Introduction.
I do this by analyzing a specific simple example of 
new Relativity postulates. A more general analysis, exploring 
a wider class of consistent Relativity postulates,
will be reported elsewhere~\cite{inprep}.

%The illustrative example here considered
%appears to be satisfactory from an experimental viewpoint,
%since it differs from ordinary Special Relativity only for high-energy
%processes, processes which of course did not form Einstein's intuition
%about Relativity. The proposal is also satisfactory from the 
%(less reliable) perspective of conceptual and aesthetic criteria;
%in fact, it turns out that the small revision of the Relativity 
%postulates which I consider does not lead to a ``symmetry loss"
%(in algebraic language: there are as many symmetry generators
%as in the usual special-relativistic case)
%and preserves unchanged the Relativity
%Principle: the laws of physics are the same for all (inertial) observers.

In the next Section I propose a general type of Relativity postulates
with observer-independent scales of both velocity and length,
and formulate a specific illustrative example of such postulates
to be analyzed in the rest of the paper.
In Section 3 I discuss the rules that relate the momentum and energy
of a particle as seen by different observers according to the chosen
illustrative example of Relativity with two observer-independent scales.
In Section 4 I present the corresponding analysis for 
the conservation rules that govern scattering processes.
Section 5 is devoted to phenomenological implications.
In Section 6 I discuss the possible role of quantum algebras
in Relativity with two observer-independent scales.
A brief summary of the results obtained and a list of
open issues are given in Section 7.

\section{Relativity and observer-independent scales}

As we explore the possibility of attributing to $L_p$
a role in space-time structure on which every inertial observer agrees, 
it is useful
to describe the step from Galilean Relativity to Special Relativity
as a solution of the problem of attributing to $c$, the speed-of-light
constant, a value on which every inertial observer agrees.
In describing the new
issues raised by a possible role of $L_p$ it is natural to view
Galilean Relativity as an analysis of the Relativity Principle
based on the assumption that there would be no fundamental scales
of velocity or length. Einstein's Relativity describes the implications
of Galilei's Relativity Principle for the case in which there is
a fundamental velocity scale. My task will then be naturally
described as a study of the implications of Galilei's Relativity Principle
for the case in which one has both a fundamental velocity scale and
a fundamental length scale.

The Relativity Principle introduced by Galilei
can be simply summarized\footnote{While here
I am intentionally not planning to be historically accurate (and not
attempting to state postulates in an absolutely rigorous way) I should
briefly mention that the Relativity Principle has been much extended from 
the times of Galilei, particularly with Einstein's realization that
this principle should apply also to electromagnetic phenomena.}
as follows:
\begin{itemize}
\item{(R.P.):} The laws of physics take the same form in all inertial 
frames ({\it i.e.} these laws are the same for all inertial observers).
\end{itemize}
It is easy to see that the implications of this principle
for geometry and kinematics depend very strongly on whether there are
fundamental scales of velocity and/or length. In fact, the introduction
of a fundamental scale is itself a physical law, and therefore the
Relativity Principle allows the introduction of such fundamental
scales only if the rules that relate the observations performed
by different inertial observers are structured in such a way that
all inertial observers can agree on the value and physical interpretation
of the fundamental scales. The Galilean rules of transformation
between observers can be easily obtained by combining the Relativity
principle with the assumption (which from the perspective here adopted
could be described as the ``first Galilean law of physics")
that there are no fundamental
scales for velocity or length:
\begin{itemize}
\item{(G.L.1):} The laws of physics do not involve any
fundamental scales of velocity and/or length.
\end{itemize}
Using this postulate and the Relativity Principle one can easily
obtain all of the familiar predictions of Galilean Relativity. 
For example, the simplicity of the law $v'=v_0+v$ of composition of velocities
(the law describing the velocity measured by one observer for a projectile
which is measured to have velocity $v$ by a second observer, which is itself 
moving with velocity $v_0$ with respect to the first observer) 
can be easily derived using the fact that, 
in absence of a fundamental velocity scale,
$v'$ could only depend on $v_0$ and $v$, $v'=f(v_0,v)$,
and imposing some obvious logical constraints, such as dimensional analysis
and the properties of $f$ that follow from its definition:
$f(0,v)=v$, $f(v_0,0)=v_0$, $f(v,v_0) = f(v_0,v)$, $f(-v_0,-v) = -f(v_0,v)$.

The task accomplished by Einstein with his Special Relativity
is the one of introducing a fundamental velocity scale
in a way that would be consistent with the Relativity Principle.
In any inertial frame, the velocity of light is $c$ whether 
the light is emitted by a body at rest or by a body in 
motion. Consistently with the perspective I am adopting, let me
write this postulate as two parts of a physical law:
\begin{itemize}
\item{(E.L.1):} The laws of physics 
do not involve any
fundamental scale of length but they do involve
a fundamental velocity scale $c$.
\item{(E.L.1b):} The value of the fundamental velocity scale $c$
can be measured by each inertial observer as the speed of light.
\end{itemize}
Notice that the structure of (E.L.1b) is not unrelated with (R.P.) and
(E.L.1). In fact, one could have naively
expected that a more careful description would be required for the 
measurement procedure that specifies the physical interpretation of $c$.
For example, one could have imagined the speed of light to depend on
the velocity of the body emitting the light and on the wavelength
of the light. However, the specification of 
a wavelength dependence would have required
a reference fundamental scale of length, in conflict with (E.L.1),
while a dependence on the velocity of the emitting body
would have been in conflict with the role of $c$ as a fundamental
scale on which all inertial observers agree, in the sense of (R.P.).

From (R.P.),(E.L.1), and (E.L.1b) one can derive
the rules that relate observations performed by different inertial
observers, which turn out to be given by the Lorentz
transformation rules. The most significant differences between
Galilean-Relativity physics and Special-Relativity physics are very
directly connected with the new fundamental scale $c$. For example,
the mentioned simple Galilean velocity law $v'=f(v_0,v)=v_0+v$ would 
have not been consistent with the role of $c$ in (E.L.1) and (E.L.1b),
and it turned out to be necessary to replace it with the,
apparently less simple, law $v'=f(v_0,v;c)=(v_0+v)/(1+v_0v/c^2)$.
The transition from Galilean-Relativity physics to Special-Relativity 
physics also required that we give up the intuitive (but actually
quite cumbersome) concept of absolute time, which would have been
in conflict with the fact that the exchange of information between
two clocks with some relative velocity 
is severely constrained by (E.L.1) and (E.L.1b).

Within the perspective adopted in this Section it is natural
to describe the task of developing a relativistic 
theory\footnote{In the literature the expression ``relativistic theory"
is used to describe two closely related but distinct concepts. 
Let me clarify the difference within the specific example
of Special Relativity (Relativity with a single observer-independent
scale $c$). 
There one can describe a theory as ``relativistic" if it is a theory
of the relation between the results of measurements performed
by different inertial observers, when the theory states that this
relation is governed by Lorentz transformations.
One also describes a theory as ``relativistic" if it is a theory
of the physical laws that a single observer writes down in its
own reference frame, when the theory has the type of rotation-like
invariance associated with the Lorentz group.
These two clearly distinct types of theories ended up sharing
the denomination ``relativistic" because of the connection that exists
between them as a result of the Relativity Principle.
In this paper I use the expression ``relativistic theory" exclusively
for theories describing the relation between the results of measurements 
performed by different inertial observers in a way that is consistent
with the Relativity Principle. The (closely-related) second type of issues, 
the ones concerning the constraints imposed by the postulates on the 
laws of physics that a given observer can write down,
is postponed to future studies.}
in which the Planck length plays a role in
the fundamental structure of space-time, 
as something that requires us to revise (E.L.1):
\begin{itemize}
\item{(L.1):} The laws of physics 
involve a fundamental velocity scale $c$
and a fundamental length scale $L_p$.
\end{itemize}
In order to complete the new relativistic theory 
I should also describe the measurement procedures that
provide the physical interpretation of $c$ and $L_p$, {\it i.e.}
I should provide a law (L.1b) 
that plays the role of (E.L.1b) in the
new relativistic theory and add an extra law (L.1c) which 
provides the physical interpretation of $L_p$.
There is no reason for the presence of the Planck length 
to require a serious revision of (E.L.1b): we should only be
careful with the fact that the presence of such a reference length
scale does raise the possibility that the speed of light might
have some wavelength dependence. Since Michelson-Morley (and 
the century of additional tests that followed) only dealt with
light with wavelengths much larger than $L_p$,
I shall address this possible "risk" of a wavelength dependence
of the speed-of-light by replacing (E.L.1b) with the following more
prudent\footnote{Notice that (L.1b) is not really conceptually 
different from (E.L.1b):
Einstein could have adopted (L.1b) and derive the wavelength
independence from the absence of a fundamental length scale
postulated in (E.L.1).
The fact that I replace (E.L.1b) with (L.1b) should not be interpreted
as suggesting that the speed of light should necessarily have
some wavelength dependence in a relativistic theory in which there is
also an observer-independent length scale. The presence of an 
observer-independent length scale renders such a wavelength dependence
possible, thereby requiring a more prudent formulation of the definition
of $c$, but there might well exist consistent
relativistic scenarios in which an observer-independent length scale
is introduced in such a way that the speed of light remains 
wavelength independent.}
statement 
\begin{itemize}
\item{(L.1b):} The value of the fundamental velocity scale $c$
can be measured by each inertial observer as the speed of light
with wavelength $\lambda$ 
much larger than $L_p$ (more rigorously, $c$ is obtained as 
the $\lambda/L_p \rightarrow \infty$ limit of the speed of light).
\end{itemize}

For the final element of the new relativistic theory, a 
postulate (L.1c) describing the role of $L_p$ in space-time structure and
kinematics, we do not have enough experimental 
information to make an educated guess. There are many physical arguments 
and theoretical models that predict one or another physical role for the 
Planck length, but none of these scenarios has any experimental support. It 
is still plausible that the Planck length has no role 
in space-time structure and kinematics 
(which would render the present analysis purely academic). The fact that 
combining some other physical scales we can construct this quantity $L_p$ 
with dimensions of a length does not prove or even suggest in itself that 
$L_p$ should be a physically meaningful length scale. And even if it is, as 
long as we do not have experimental information on its physical 
interpretation we can only speculate that this interpretation be given by 
one of the many scenarios already being discussed or by some other scenario 
yet to be conceived. When the Relativity Principle was made agree with the 
presence of the fundamental velocity scale $c$, there was already robust 
data suggesting a physical interpretation of $c$. I am now exploring the 
possibility that the Relativity Principle may coexist with both a 
fundamental velocity scale $c$ and a fundamental length scale $L_p$ at a 
time when we do not (yet) have any robust hints on the physical 
interpretation of $L_p$. However, I am here only arguing that, from the 
point of view of the formal exercise of developing a relativistic theory in 
which the Relativity Principle (R.P.) coexists with fundamental scales, the 
introduction of the scale $L_p$ is genuinely not very different from the 
introduction of the scale $c$, 
and I can provide support for my argument by showing 
that there are at least some examples of the postulate (L.1c) which lead to 
a consistent conceptual framework. The actual form that (L.1c) takes in 
Nature is likely to be different from any proposal we can contemplate 
presently, while we are still lacking the needed guidance of experimental 
information, but through the study of some specific examples we can already 
acquire some familiarity with the new elements required by a conceptual 
framework in which the Relativity Principle coexists with 
observer-independent
scales of velocity and length. 

In light of this situation the goal of the present analysis is not
the one of obtaining the correct relativistic theory with fundamental
scales of velocity and length, but just to show that such a theory
can be logically consistent. The hope is that this analysis will
provide encouragement to experimentalists\footnote{In a sense, in the
case of Special Relativity experimental information came first and
the proof of logical consistency came with the work of Einstein,
while I am here proving the logical consistency first in the
hope that this may provide additional encouragement for future
experimental searches of a physical role for $L_p$.}
now starting to be involved
in searches of a role for $L_p$ in space-time structure. 
Consistently with this motivation, in remainder of this paper I
analyse the implications of one specific illustrative example
of the postulate (L.1c):
\begin{itemize}
\item{(L.1c$^*$):} Each inertial observer can establish the value 
of ${\tilde L}_p$ (same value for all inertial observers)
by determining the dispersion relation for photons, which takes
the form $E^2 - c^2 p^2 + f(E,p;{\tilde L}_p)=0$, where 
the function $f$ is the same for all inertial observers
and in particular all inertial observers agree on the
leading ${\tilde L}_p$ dependence 
of $f$: $f(E,p;{\tilde L}_p) \simeq {\tilde L}_p c p^2 E$
\end{itemize}
The ``$^*$" in (L.1c$^*$) is a reminder of the fact that this 
postulate is adopted here only for the explorative objectives
described above.
(L.1c$^*$) involves ${\tilde L}_p \equiv \rho L_p$ 
(with $\rho \in R$)
rather than directly $L_p$ 
in order to leave room 
for a numerical coefficient and a possible sign choice
%({\it e.g.} ${\tilde L}_p = {\pm} 4 \pi^2 L_p$)
between the quantity
setting the strength\footnote{As illustrated
by the specific example (L.1c$^*$), the observer-independent
length scale must not necessarily have the physical meaning
of the length of something. For example, as illustrated by (L.1c$^*$),
the role of $L_p$ in space-time structure could be such that
it provides a sort of reference scale for wavelengths.} 
of the dispersion-relation
deformation and the Planck length calculated {\it a la} Planck.

As I shall emphasize in the following, and I already
briefly mentioned in the Introduction,
this particular example of the postulate (L.1c) is rather interesting
in light of the role it could play in 
experiments~\cite{grbgac,schaef,billetal} on 
wavelength dependence of the speed of light
and in experiments~\cite{kifu,aus,ita,gactp1,kluz,sato} 
attempting to determine the threshold
energies needed for certain particle-physics processes.
%Also notice that if (L.1c$^*$) turned
%out to be the correct version of the (L.1c) law giving
%the physical interpretation of $L_p$,
%then (L.1b) would be redundant: the dispersion relation
%adopted in (L.1c$^*$) could be used to determine both $c$ and $L_p$.

In the remainder of this paper I shall explore the logical
consistency of a relativistic theory based 
on (R.P.), (L.1), (L.1b) and (L.1c$^*$).

\section{Transformation rules (one-particle case)}

The logical consistency of the 
postulates (R.P.), (L.1), (L.1b), and (L.1c$^*$) requires
that, in their analyses of photon data 
in leading order in ${\tilde L}_p$, 
all inertial observers agree on
the dispersion relation $E^2 = c^2 p^2 - {\tilde L}_p c p^2 E$,
for fixed (observer-independent) values of $c$ and ${\tilde L}_p$.
The postulates do not explicitly concern massive particles,
which are at rest ($p=0$) in some inertial frames and in those frames
have a ``rest energy" which we indentify with the mass $m$.
For massive particles I adopt the
dispersion relation $E^2 = c^4 m^2 + c^2 p^2 - {\tilde L}_p c p^2 E$,
which satisfies these properties.
I postpone to future studies~\cite{inprep} the possibility that the
postulate (L.1c$^*$) might coexist with more complicated dispersion
relations for massive particles of the 
type $E^2 = c^4 m^2 + c^2 p^2 - {\tilde L}_p c p^2 E + F(p,E;m;{\tilde L}_p)$,
which are consistent with (L.1c$^*$) not only in the case $F=0$
(here considered) but also whenever $F$ is such 
that $F(p,E;0;{\tilde L}_p)=F(p,E;m;0)=F(0,E;m;{\tilde L}_p)=0$.

In the undeformed Lorentz case the transformations between
different inertial frames are
characterized by the familiar differential generators
of rotations ($R_a$) and boosts ($B_a$):
\begin{equation}
R_a = - i \epsilon_{abc} p_b {\partial \over \partial p_c} ~,
\label{rotnormal}
\end{equation}
\begin{equation}
B_a = i c p_a {\partial \over \partial E} 
+ i c^{-1} E {\partial \over \partial p_a}
~.
\label{boosnormal}
\end{equation}
Clearly the consistency of the postulates 
(R.P.), (L.1), (L.1b), and (L.1c$^*$) does not require
any deformation of rotations. But boosts must clearly be deformed,
and this is after all what we should have expected in light of the
points reported in the Introduction.\footnote{Moreover, a deformation
of boosts is roughly a sign that the time direction is somewhat
different from the space directions, something foreign
to ordinary Special Relativity, but intuitively consistent with the nature
of most of our observations.} 
In this exploratory study
I shall simply focus on boosts along the direction of motion
of the particle (which are the most troublesome)
and, consistently with the prudent formulation of the exploratory
postulate (L.1c$^*$), I shall be satisfied with verifying consistency at 
leading order in ${\tilde L}_p$.

Let us therefore consider a particle moving along the $z$ direction
with momentum $p_z$, mass $m$, and 
energy such that $E^2= c^2 p_z^2 + c^4 m^2 - {\tilde L}_p c p_z^2 E$. 
According to a second
inertial observer the same particle has
momentum and energy $p_z',E'$.
This second observer, as seen by the first observer,
is moving along the $z$ axis.
The relation between $p_z',E'$ and $p_z,E$ must be consistent
with the postulates (R.P.), (L.1), (L.1b), and (L.1c$^*$). 
I make the following ansatz for the deformed $z$-direction boost
\begin{equation}
B_z^{{\tilde L}_p} 
= i [c p_z + {\tilde L}_p \Delta_1(p_z^2,E)] {\partial \over \partial E} + i 
[E/c + {\tilde L}_p \Delta_2(p_z^2,E)] {\partial \over \partial p_z}
~,
\label{boosdelta}
\end{equation}
where $\Delta_1$ and $\Delta_2$ (on which I impose the property of depending 
on $p_z$ only through $p_z^2$ with the objective of working consistently with
overall rotational invariance)
are deformation functions to be determined
by consistency with the postulates.
This consistency basically demands that (rotations and) boosts
leave the dispersion relation unaffected (all inertial observers must agree
on the dispersion relation). Imposing this condition
the form of $\Delta_1$ and $\Delta_2$
is easily determined, and the $z$-boost generator takes the form
%\footnote{It is perhaps here worth reminding again to the reader that all 
%formulas are being analyzed and reported in leading order in ${\tilde L}_p$.}
\begin{equation}
B_z^{{\tilde L}_p} 
= i c p_z {\partial \over \partial E} + i 
[E/c + {\tilde L}_p E^2/c^2 
+ {\tilde L}_p p_z^2/2] {\partial \over \partial p_z}
~.
\label{boosnew}
\end{equation}

This form of the $z$-boost already assures the logical consistency
of the observations of our particle moving along the $z$ axis
as reported by our two inertial observers.
Both observers agree that in their own reference frame
the law of propagation is the one described by postulate (L.1c$^*$),
and from the analysis of data on this propagation
they both extract the same values of $c$ and ${\tilde L}_p$.
Using the Relativity Principle (R.P.) they can consistently interpret
their different observations for what concerns the values
of the momentum and energy of the particle (these values are
of course different, but they are, as imposed by the new postulates, 
related by the boost transformations described by (\ref{boosnew})).

It is useful to obtain explicit formulas
for the finite $z$-boost transformations that relate
the observations of our two observers. By only exhibiting
the $z$-boost generator I have only described
infinitesimal transformations:
\begin{equation}
{dE \over d\xi} = i [B_z^{{\tilde L}_p},E] = - c p_z ~,
\label{infe}
\end{equation}
\begin{equation}
{dp_z \over d\xi} = i [B_z^{{\tilde L}_p},p_z] = 
-E/c - {\tilde L}_p E^2/c^2 - {\tilde L}_p p_z^2/2
~.
\label{infp}
\end{equation}
In spite of the reacher structure of the new $z$-boost generator,
the derivation of finite transformations from the structure
of the generator of infinitesimal transformations is not
significantly more complicated than in the Lorentz case.
With simple, but somewhat tedious, calculations one finds that
\begin{equation}
E(\xi) = - c^{-1} \alpha e^\xi + c^{-1} \beta e^{-\xi} 
+ {\tilde L}_p c^{-2} {\alpha^2 \over 2} e^{2 \xi} 
+ {\tilde L}_p c^{-2} {\beta^2 \over 2} e^{- 2 \xi}
+ {\tilde L}_p c^{-2} \alpha \beta ~,
\label{finitee}
\end{equation}
\begin{equation}
p_z(\xi) = \alpha e^\xi + \beta e^{-\xi} 
- {\tilde L}_p \alpha^2 e^{2 \xi} 
+ {\tilde L}_p \beta^2 e^{- 2 \xi}
~.
\label{finitep}
\end{equation}
where $\alpha$ and $\beta$ depend on $E_0 \equiv [E]_{\xi = 0}$ 
and $p_{z,0} \equiv [p_z]_{\xi = 0}$:
\begin{equation}
\alpha = c p_{z,0}/2 - E_0/2 - {\tilde L}_p E_0 p_{z,0}/2 
+ {\tilde L}_p c p_{z,0}^2/4  ~, 
\label{aformula}
\end{equation}
\begin{equation}
\beta = c p_{z,0}/2 + E_0/2 - {\tilde L}_p E_0 p_{z,0}/2 
- {\tilde L}_p c p_{z,0}^2/4 
~.
\label{bformula}
\end{equation}
Of course, these solutions $E(\xi),p_z(\xi)$ are such 
that $E(\xi)^2 - c^2 p_z(\xi)^2 - c^4 m^2 
+ {\tilde L}_p c p_z(\xi)^2 E(\xi) =0$. 
Also reassuringly, the limit ${\tilde L}_p \rightarrow 0$ of the analysis
reproduces ordinary Lorentz $z$-boost transformations.

Having derived the transformation rules for energy and momentum,
in the analysis of a classical space-time framework this would be
the right point for a discussion of the corresponding transformation rules
for space-time coordinates. However, the intuition which guides the 
analysis here being reported is that space-time (even flat space-times)
should be in one way or another quantized (if indeed the Planck length
plays a role in its structure, as here assumed), and in that case the issue
of imposing the Relativity Principle on coordinate transformations
appears somewhat delicate.
The Relativity principle essentially concerns the observations made
by different inertial observers, but if (for example) the proper
formal tool for the description 
of coordinates in quantum space-time is represented by
operators (rather than classical variables) one might obtain
missleading results by formally insisting on a sort of Relativity Principle
for transformations of these quantum coordinates. 
%It is reasonable to expect that
%the actual coordinates assigned by an observer are still classical, even when
%the supporting mathematics is the one of quantum coordinates (just like
%in ordinary Quantum Mechanics the measurements performed 
%by an observer always attribute to the 
%spin of a particle a number, even though in the supporting mathematics
%the spin is associated with an operator). 
%The relation between the (conjectured) quantization of space-time and the 
%transformations rules valid for the classical coordinates operatively
%attributed by an observer to an event appears to be very subtle,
%and it may affect in a non-trivial way the (usually straightforward)
%connection between energy-momentum transformation rules 
%and space-time transformation rules. 
It appears prudent, at this early
stage of analysis of Relativity with observer-independent $c$ and $L_p$,
to postpone this delicate issue.

\section{Kinematical conditions for processes (two-particle case)}

In the preceding Section I verified the logical consistency of the 
postulates (R.P.), (L.1), (L.1b), and (L.1c$^*$) in the one-particle
sector. As they have been stated in Section 2, the new postulates
explicitly refer only to the one-particle sector.
Some ideas on the structure of space-time at Planckian 
distances might motivate the study of otherwise unexpected
sharp differences between the one-particle 
and the two-particle sector. For example, such sharp differences
appear to be likely in scenarios in which particles are described
as geometry ``defects".
Again I shall postpone the analysis of such more exotic possibilities
to studies~\cite{inprep} that will follow the present exploratory 
analysis. Here I just focus on one aspect of the two-particle
sector: the laws of conservation that characterize $2 \rightarrow 2$
scattering (collision processes with two incoming particles
and two, possibly different, outgoing particles).
This exploratory analysis of collision processes is sufficient
to clarify that in relativistic theories with an observer-independent
length scale some familiar assumptions must be given up, but
one can maintain the requirements pertaining to the objectivity
of physical processes.

If indeed the hypothesis here being explored, that the Planck length is
an observer-independent physical scale of space-time structure, 
turns out to be verified
in Nature, we will certainly be forced to revise many
familiar concepts. By adding a second fundamental scale in Relativity
we should inevitably encounter concepts that are profoundly new,
in the same sense that the introduction of 
the first fundamental scale, $c$, led to new concepts such as 
the relative time and the mentioned deformation of the Galilean
law of composition of velocities. In the present exploratory study,
which is lacking the important guidance of experimental information
on Planck-length physics, I am of course not attempting to formulate
a general description of such new concepts. On the contrary I would like
to identify at least a few concepts which should not
require modification, even in presence of the drastically new
ingredient of an observer-independent length scale.
One of such "nonnegotiable" aspects of Relativity is the fact
that two observers, even when they are in relative motion,
should agree on the occurrence of physical processes.
It is perhaps worth stating explicitly a condition (usually implicit
in Relativity postulates) that this objectivity of physical
processes imposes on the rules of kinematics
\begin{itemize}
\item{(R.P.addendum):} The conservation laws 
that must be satisfied by physical processes should be covariant
under the transformations that relate the kinematic properties
of particles as measured by different observers
({\it i.e.} all observers should agree on whether or not
a certain process is allowed).
\end{itemize}

This addendum imposes important constraints on the conservation laws.
The constraints are trivially satisfied in ordinary Special Relativity.
Let me discuss this in the simple case 
of a scattering process $a+b \rightarrow c+d$ (collision
processes with incoming particles $a$ and $b$ and outgoing
particles $c$ and $d$). 
Also in this Section my considerations are essentially 
one-dimensional.
In three space dimensions one-dimensional kinematics is relevant
for head-on $a$-$b$ collisions producing $c$-$d$ at threshold
(when the kinematical conditions are only barely satisfied and therefore
the particles produced do not have any energy available for momentum 
components in the directions orthogonal to the one of the head-on
collision). Collisions at threshold are after all the most interesting
collisions,
since they force us to insist on the fact that all inertial observers agree
when the delicate kinematic balance of threshold production is realized.
The special-relativistic kinematic requirements for 
such processes are $E_a + E_b - E_c -E_d=0$ and $p_a + p_b - p_c -p_d=0$.
Using the special-relativitic transformation rules, $d E_j/d\xi = - p_j$,
$dp_j/d\xi = -E_j$, one immediately verifies that when the requirements
are satisfied in one inertial frame they are also verified in all
other inertial frames:
\begin{equation}
{d (E_a + E_b - E_c -E_d) \over d \xi} = 
-p_a - p_b + p_c +p_d
~,
\label{conservnormale}
\end{equation}
\begin{equation}
{d (p_a + p_b - p_c -p_d) \over d \xi} = 
-E_a - E_b + E_c +E_d
~.
\label{conservnormalp}
\end{equation}

The requirements $E_a + E_b - E_c -E_d=0$ 
and $p_a + p_b - p_c -p_d=0$ 
cannot be imposed in the new relativistic framework
which I am analyzing. Because of the structure of the transformation
rules $d E_j/d\xi = - c p_j$, 
$dp_j/d\xi = -E_j/c - {\tilde L}_p E_j^2/c^2 - {\tilde L}_p p_j^2/2$
the requirements $E_a + E_b - E_c -E_d=0$ 
and $p_a + p_b - p_c -p_d=0$ would not satisfy the (R.P.addendum).
In order to satisfy the (R.P.addendum) it appears sufficient
to replace the requirements $E_a + E_b - E_c -E_d=0$, 
$p_a + p_b - p_c -p_d=0$ with the requirements
\begin{equation}
E_a + E_b + {\tilde L}_p c p_a p_b
- E_c -E_d -{\tilde L}_p c p_c p_d = 0
~,
\label{conservnewe}
\end{equation}
\begin{equation}
p_a + p_b + {\tilde L}_p (E_a p_b + E_b p_a)/c 
- p_c -p_d - {\tilde L}_p (E_c p_d + E_d p_c)/c = 0
~.
\label{conservnewp}
\end{equation}

In fact, using $d E_j/d\xi = - c p_j$ and
$dp_j/d\xi = -E_j/c - {\tilde L}_p E_j^2/c^2 - {\tilde L}_p p_j^2/2$
one easily finds that 
the requirements 
(\ref{conservnewe}),(\ref{conservnewp})
are satisfied (again, to leading order in ${\tilde L}_p$)
in every inertial frame if they are satisfied in one inertial frame.
The difference between(\ref{conservnewe}),(\ref{conservnewp})
and the ordinary requirements $E_a + E_b - E_c -E_d=0$, 
$p_a + p_b - p_c -p_d=0$ does not appear more surprising than
the difference between the Galilean velocity law, mentioned
in Section~2, and its special-relativistic version.
Both deformations are directly associated with the introduction
of the corresponding observer-independent scale.

It is also worth emphasizing that, to leading order in ${\tilde L}_p$,
the requirements (\ref{conservnewe}),(\ref{conservnewp})
are equivalent to the requirements 
\begin{equation}
E_a \! + \! E_b \! + \! {\tilde L}_p c p_a p_b \! 
+ \! \eta {\tilde L}_p c (p_a + p_b)^2 \! 
- \! E_c \! - \! E_d \! - \! {\tilde L}_p c p_c p_d \! 
- \! \eta {\tilde L}_p c (p_c + p_d)^2 \! = \! 0
~,
\label{conservnewebis}
\end{equation}
\begin{equation}
p_a \! + \! p_b \! + \! {\tilde L}_p (E_a p_b + E_b p_a)/c \! 
+ \! \zeta {\tilde L}_p (p_a + p_b)^2 \! 
- \! p_c \! - \! p_d \! - \! {\tilde L}_p (E_c p_d + E_d p_c)/c \! 
- \! \zeta {\tilde L}_p (p_c + p_d)^2 \! = \! 0
~,
\label{conservnewpbis}
\end{equation}
with $\eta$ and $\zeta$ some fixed numbers.
The fact that the presence of the observer-independent length scale
allows the formulation of two equivalent sets of kinematic requirements
appears to be potentially relevant for the identification of the total
energy and total momentum
of a two-particle system. 
One could describe (\ref{conservnewe}),(\ref{conservnewp}) as conditions
for the conservation of total energy and total momentum, but
in presence of the observer-independent length scale
it appears even conceivable (exploting the freedom 
encoded in (\ref{conservnewebis}) and (\ref{conservnewpbis})) 
to develop a thoretical framework in which the 
kinematic requirements 
are not interpreted strictly as conservation of total energy
and total momentum ({\it e.g.} total energy and total momentum
are exactly conserved only in processes in which the total
number of fundamental particles is conserved, while in other cases
there would be small deviations from energy-momentum conservation
depending on the smallness of the Planck length and, say, the number
of fundamental particles gained/lost in the process).
This issue of total energy and total momentum and their 
conservation clearly becomes more subtle in presence
of an observer-independent length scale, but it is not alarming
with respect to the logical consistency of Relativity with
two observer-independent scales: we are just confronted with the risk 
of having more than one logically consistent scenario 
(so the problem might be the one of choosing among more than one 
consistent possibility, rather than the
one of not having a consistent solution).
I therefore postpone it to future studies.

\section{Phenomenology}

The postulate (L.1c$^*$), on which I based the present exploratory 
study of the consistency of Relativity with two observer-independent
scales, involves a deformation of the dispersion relation
whose phenomenological implications have already been considered 
in previous studies (see, {\it e.g.}, 
Refs.~\cite{grbgac,kifu,aus,schaef,billetal,ita,gactp1,kluz,sato,gactp2}),
but those previous studies made the (explicit or implicit)
assumption that the deformation would reflect a Relativity violation, 
of the type one would expect in presence
of background fields or a medium.\footnote{Here a simple relevant
analogy is the one the familiar
special-relativistic description of the motion of an electron
in the background of electromagnetic fields. 
This physical context is described by different 
observers in a way that is consistent
with the Relativity Principle, but only when these
observers take into account
the fact that the background electromagnetic fields
also take different values in different inertial frames.
Observers assuming that there are no background
electromagnetic fields  
would deduce from their observations that the Relativity Principle
is violated (effective theories written down without considering
the background would formally violate Lorentz invariance even in
a special-relativistic framework).
Within Special Relativity (with its single observer-independent
scale $c$)
the Planck length could of course be introduced~\cite{grbgac} 
in space-time
physics in association with an accompanying background
(something like a ``quantum-gravity medium"),
but this could be used to single out a preferred class
of inertial observers for the description of space-time structure.
This again is analogous to the mentioned context in electromagnetism:
the electric and the magnetic components of the background
are not observer-independent, and could be used to single out
a class of inertial observers.
In the present paper I am observing that it is possible
to introduce the Planck length in a way that does not
lead to a preferred class of inertial observers, and I reserve
the description ``relativistic short-distance structure 
of space-time" to space-times in which the Planck length
has this type of role. When the Planck length is introduced
in space-time physics in a way that allows the selection
of a preferred class of inertial observers I use the 
description ``non-relativistic short-distance 
structure of space-time".
My results show that there are
at least two ways for Nature to make use of the Planck length
in space-time physis:
the scenario previously considered~\cite{grbgac} in which
Special Relativity is not modified, but the Planck length is
introduced together with an accompanying background (``medium")
and an accompanying class of preferred inertial observers,
and the new scenario here proposed in which the Planck length 
is introduced without an accompanying background (and the
associated preferred class of inertial observers), 
but at the cost of a revision of Special Relativity
in which the Planck length acquires the status of 
an observer-independent property of space-time.}
The fact that here I have explored the possibility that
the same dispersion relation be accomodated in a theoretical
framework which is still genuinely relativistic (with the only
new ingredient of having a second observer-independent physical scale)
changes the perspective of these phenomenological studies.
I find that also in the relativistic framework the two signatures
which have attracted most attention, wavelength dependence of
the speed of light and threshold anomalies, are present,
but, of course, some features are modified by the requirement
that the same dispersion relation is valid 
in all inertial frames.

\subsection{Wavelength-dependent speed of light}

Both in the relativistic framework here proposed and in the
non-relativistic schemes considered in other studies
it is natural to assume that the dispersion 
relation $E^2 = c^2 p^2 - {\tilde L}_p c p^2 E$
would imply a wavelength dependence of the speed of light
described by the formula
\begin{equation}
v = c \, (1 - {\tilde L}_p c^{-1} E/2)
~.
\label{eqvelocity}
\end{equation}
However, in the present relativistic framework ${\tilde L}_p$
takes the same value in all inertial frames, contrary to the case
in which (\ref{eqvelocity}) is a manifestation of 
the existence of a preferred class of inertial observers.
This clearly should give rise to different expectations for
the spectrum and time-of-arrival history of the gamma rays
we detect from catastrofic explosive events in
far away galaxies.
A detailed phenomenological analysis is postponed to future studies,
but it appears that
a key tool for distinguishing between the two scenarios
could be provided by the fact that different
galaxies have different velocities as seen from the
frame in which our detectors are at rest.

Setting aside the issue of the differences between relativistic
and non-relativistic schemes, the central issue remains the fact
that some wavelength dependence should be seen in order to provide
support for (\ref{eqvelocity}). All available data are consistent
with wavelength independence, but this is not surprising
since only detectors to be operative
in a few years will~\cite{grbgac,schaef,billetal}
reach sensitivity levels sufficient for the detection
of the small wavelength dependence predicted by (\ref{eqvelocity}).
The GLAST collaboration already has~\cite{glast} on its agenda
searches of this wavelength dependence.
The fact that the sought effect can 
also be accommodated in a framework
that does not require the short-distance structure of space-time
to select a preferred class of inertial observers
should provide
an additional source of motivation for these planned studies.

\subsection{Threshold anomalies}

As emphasized in Section~4,
an important aspect of Special Relativity is the fact that the
requirements $E_a + E_b - E_c -E_d=0$ 
and $p_a + p_b - p_c -p_d=0$ 
must be satisfied by particle-physics processes.
For example, according to these requirements
a collision between a soft photon and a high-energy photon
is kinematically allowed to
produce an electron-positron pair only if the high-energy
photon has energy $E$ greater than (or equal to) the threshold
energy $E_{threshold} = c^4 m_e^2/\epsilon$,
where $\epsilon$ is the energy of the soft photon and $m_e$
is the electron mass.
Similarly, a collision between a soft photon and a high-energy proton
can give proton+pion only if the high-energy
proton has energy $E$ greater than (or equal 
to) $E_{threshold} \simeq c^2 (m_\pi m_p/2 + m_\pi^2/4)/\epsilon$.
In addition to quantum-gravity arguments~\cite{grbgac}
and the mentioned recently-acquired capability of experimental
studies of the conjectured wavelength dependence of 
the speed of light~\cite{grbgac,schaef,billetal},
part of the recent interest in deformed/anomalous dispersion relations
has been motivated~\cite{kifu,aus,ita,gactp1,colgla,kluz,sato,gactp2}
by puzzling astrophysics data which admit interpretation as
manifestations of deviations from the special-relativistic
threshold conditions for electron-positron pair production
and photopion production.
These puzzling data could indicate that the threshold energies
calculated using Special Relativity are too low.

Several studies~\cite{kifu,aus,ita,gactp1,kluz,sato,gactp2}
have observed that the deformed dispersion 
relation $E^2 = c^4 m^2 + c^2 p^2 - {\tilde L}_p c p^2 E$
when combined with the conditions $E_a + E_b - E_c -E_d=0$ 
and $p_a + p_b - p_c -p_d=0$ leads to a large shift
of the relevant astrophysics thresholds. These threshold
shifts are of order ${\tilde L}_p c^{-1} E^3/\epsilon$ (and in the
relevant contexts ${\tilde L}_p c^{-1} E^3/\epsilon > E$)
and would be sufficient to solve the observed threshold paradoxes.

While these previous analyses are of course extremely interesting,
the fact that they require 
the short-distance structure of space-time
to select a preferred class of inertial observers
can be unwelcome.
Since I am proposing a conceptual framework in which
deformed dispersion relations can coexist with 
a relativistic description of the short-distance structure 
of space-time,
it is interesting to check whether the relativistic treatment
of the dispersion 
relation $E^2 = c^4 m^2 + c^2 p^2 - {\tilde L}_p c p^2 E$
does lead to threshold anomalies.
It is easy to verify that 
the results obtained in Section~4 do imply threshold anomalies
and that for positive ${\tilde L}_p$
these threshold anomalies consistently 
go in the right direction to explain the puzzling data
(they correspond to an upward shift of the threshold).
However, basically
as a result of the strict
constraints imposed by the Relativity Principle,
the threshold shift is much smaller, only of order ${\tilde L}_p E^2/c$,
and would not be sufficient to solve the paradoxes.
Of course, within the relativistic framework here developed,
with two observer-independent physical scales,
the postulate (L.1c$^*$), which involves the dispersion 
relation $E^2 = c^4 m^2 + c^2 p^2 - {\tilde L}_p c p^2 E$,
is only an illustrative example.
The only general point on threshold anomalies that can be extracted
from the present study is that threshold anomalies can coexist
with a relativistic description of the short-distance structure 
of space-time,
but the magnitude of the anomalies
will depend strongly on the correct form of the postulate (L.1c)
which is realized in Nature.
A threshold anomaly is NOT necessarily 
a signature of space-time structure that selects preferred
inertial observers, but only a signature of departure from
a strictly special-relativistic analysis
(with its single observer-independent scale $c$).
If the preliminary data providing evidence for these threshold paradoxes
are confirmed by more refined experiments, it will then be necessary
to find ways to check experimentally whether the anomalies
are due to the dispersion 
relation $E^2 = c^4 m^2 + c^2 p^2 - {\tilde L}_p c p^2 E$
analyzed in a way that gives rise to a preferred class of inertial
observers or to some other dispersion relation analyzed according to
the new relativistic framework here proposed.
In this respect an important role could be played by the function $F$
introduced in Section~3. In fact, my result of a weak threshold
anomaly appears to depend strongly on the assumption $F=0$
in the analysis of massive particles.
With different choices of $F$ it appears possible
to obtain stronger threshold anomalies, providing a solution of 
the paradoxes. For example, if the function $F$ is such that
around $10^{20}eV$ the correction to the proton dispersion 
relation is of order ${\tilde L}_p c^{-3} E^4/m_{proton}$
or ${\tilde L}_p^2 c^{-5} E^6/m_{proton}^2$
one should obtain a 
solution of the observed violations of the GZK threshold limit 
for cosmic rays. However, at this early stage of study of
Relativity with observer-independent $c$ and $L_p$,
it appears difficult to identify general criteria
to be satisfied by $F$ in order to achieve consistency with
the type of relativistic frameword here proposed.
Work on general criteria for the consistency of the function $F$
appears to be strongly motivated,
since it might lead to the interpretation of the
threshold paradoxes observed in astrophysics 
as the first manifestation of an
observer-independent length scale in the short-distance structure
of space-time.

\section{Relation with quantum symmetries}

It is natural to assume that the coexistence of the Relativity Principle
with length and velocity observer-independent scales should lead to
the emergence of space-time symmetries, just in the same sense that
Special Relativity, with its single observer-independent scale,
leads to Lorentz symmetry. The fact that the new symmetries should
involve an additional scale and should reproduce ordinary Lorentz invariance
in a certain limit of the additional scale (the $L_p \rightarrow 0$ limit)
suggests that the subject of ``quantum groups" and ``quantum algebras"
should be in some way relevant. 
It is probably too early to conclude that this connection should
characterize all examples of the type of relativistic theories
here proposed (theories in which the Relativity Principle coexists 
with observer-independent scales of both velocity and length), 
but it definitely characterizes the specific example I considered
here in detail.
In fact, the postulate (L.1c$^*$) involves a dispersion
relation which corresponds to the leading-order-in-${\tilde L}_p$
version of a casimir that has emerged~\cite{majrue,kpoinap} 
in the quantum-algebra literature, and, upon imposing consistency
with the Relativity Principle, I was led to boost (and rotation)
generators which can also be recognized as
the leading-order-in-${\tilde L}_p$
version of the generators of the relevant quantum algebra.
Like\footnote{It would be nice to be encouraged by the analogy
that one can make with the time
when Einstein analyzed the coexistence of the Relativity Principle
with a single observer-independent (velocity) scale: like coexistence
with a velocity scale led Einstein to preexisting mathematics of Fitzgerald
and Lorentz, here, within the chosen illustrative example, I was
led to preexisting quantum-algebra mathematics.
Of course, this amusing analogy cannot provide too much encouragement.
The most crucial source of encouragement
for a physical theory is still missing:
in 1905 there was substantial experimental evidence of an observer-independent
velocity scale, while nothing in presently-available experimental data
appears to require an observer-independent length scale.
All we have are a few (of course, debatable) theoretical arguments
suggesting that we should find room for the Planck length.}
Special Relativity describes a possible role for the Lorentz
algebra in physics, the illustrative example of Relativity with two
observer-independent physical scales here considered led me to a possible
role in physics for the quantum algebra proposed 
in Refs.~\cite{majrue,kpoinap}.

I was unable to find in the mathematics literature 
the finite boost trasformations (\ref{finitee}),(\ref{finitep}),
but, based on comparison with the analysis in Ref.~\cite{rueggnew}
(which concerned a rather similar quantum algebra), 
I am confident that the formulas (\ref{finitee}),(\ref{finitep}),
have been derived consistently with
the spirit of quantum algebras.

While the one-particle sector appears to be fully consistent with
the mathematics of quantum algebras, the analysis of two-particle
processes reported in Section~4 appears to require some new 
algebraic tools.
In particular, at least according to the standard interpretation
of the strictly mathematical language of analysis of quantum algebras,
the mathematics literature would support 
the expectation~\cite{majrue,kpoinap}
that the composition of momenta in the two-particle sector
should involve a troubling asymmetry between pairs of particles.
Even in the case of two identical particles it appears necessary to 
handle to the two momenta in a nonsymmetric way, while the composition
of energies is undeformed.
On the contrary, the analysis reported in Section~4
gives a fully symmetric description of two-particle processes,
but it appeared necessary to modify the conservation laws
of both energies and momenta.
It appears therefore plausible that a mathematical description
of the findings reported in Section~4
may require the introduction of new concepts in the subject
of quantum algebras.

Another subject which might require some mathematics work
is related to the double role of relativistic symmetries.
For example, in Special Relativity one has on the one hand
Lorentz transformations, which provide
the map between the results of measurements performed by different
inertial observables, and on the other hand, combining these
transformation properties with the Relativity Principle,
one must also impose Lorentz symmetry of the physical laws that each
observer writes down in its own frame.
A rich phenomenology should emerge from imposing a similar
double role of the transformations here considered for a
relativistic theory with observer-independent velocity and length
scales.

\section{Summary and outlook}

\subsection{Relativity can be doubly special}

From the viewpoint advocated in Section~2 the Relativity Principle
is somewhat hostile to the introduction of observer-idependent
physical scales. In that respect, Einstein's Relativity postulates
well deserve to be qualified as ``special", since they provide an example
in which the Relativity Principle coexists with an 
observer-independent (velocity) scale.
In this paper I argued that Relativity can also be ``doubly special",
in the sense that the Relativity Principle can coexist with 
observer-independent scales of both velocity and length.
I was unable (it actually was beyond the scope of this first exploratory
study~\cite{inprep}) 
to produce a list with all possible ways to introduce an
observer-independent length scale in Relativity, but I examined
one illustrative example of postulates and showed
its logical consistency.

\subsection{Phenomenology with Auger and GLAST}

I considered the particular illustrative 
example (R.P.), (L.1), (L.1b), (L.1c$^*$)
of Relativity postulates
with observer-independent velocity and length scales 
also because it shows that the issue of a possible
role of the Planck length in Relativity is not merely academic:
it can have observable consequences. In this first exploratory
study I just focused on some phenomena, wavelength dependence of the 
speed of light~\cite{grbgac} and threshold anomalies~\cite{gactp1},
which had already attracted interest in some schemes in which the 
dispersion relation $E^2 = c^2 p^2 - {\tilde L}_p c p^2 E$
is conjectured to emerge as a manifestation of a sort 
of ``quantum-gravity ether" (the so-called ``quantum-gravity vacuum").
I showed in Section~5 that within a few years,
experiments with good sensitivity to a possible wavelength dependence
of the speed of light (such as GLAST~\cite{glast})
and experiments capable of providing insight on the mentioned threshold
paradoxes (such as studies planned by 
the Pierre Auger Observatory~\cite{auger})
will test in detail
the illustrative example of new Relativity postulates
here considered, and (should new-physics
effects be found at all) the new relativistic framework
can be distinguished from 
scenarios in which the same dispersion 
relation $E^2 = c^2 p^2 - {\tilde L}_p c p^2 E$
is described in a way that would allow to select a preferred class
of inertial frames.

It is rather satisfactory that experiments on wavelength dependence
of the speed of light will test the predictions derived in Section~3,
concerning the one-particle sector, while experiments on threshold 
anomalies will test the predictions derived in Section~4,
concerning the two-particle sector. The illustrative example here 
considered is therefore going to be tested in two independent
and significantly-different ways.

\subsection{Other forms of the postulate (L.1c)}

Even assuming (and there is nothing in presently available
experimental data that would suggest it, one can only
produce some tentative theoretical arguments for it)
that Nature makes use of the possibility (here shown to be viable)
of a relativistic framework with observer-independent 
velocity and length scales, we presently have no reason
to assume that the specific postulate (L.1c$^*$), which I here used
as illustrative example of the postulate (L.1c),
is the one chosen by Nature. An important 
issue for future developments of this new relativistic framework
is the one of finding other logically consistent possibilities
for the postulate (L.1c).

In general it is not even clear that $L_p$ should be introduced
in the Relativity postulates through a deformed dispersion relation
(there are other possibilities~\cite{inprep}), but,
using (L.1c$^*$) as starting point, one could begin with the exploration
of other possible postulates by trying to formulate a postulate
that reproduces (L.1c$^*$) in leading order but provides the
precise $L_p$ dependence of the dispersion relation to all orders
in $L_p$. Such an alternative postulate could
then be studied naturally by imposing the conditions for logical
consistency of the postulates as exact relations (rather than only
as leading-order relations as done here). In light of the results on 
relevant quantum algebras~\cite{majrue,kpoinap} and the fact that,
as it emerged from the analysis of the illustrative example here considered,
there might be a connection between these quantum-algebra results
and some formulations of the (L.1c) postulate,
it appears quite
plausible that such an all-order formulation exists (but, based on the
results obtained for the illustrative example here considered, 
it appears likely that the mathematics of quantum algebras
would not play a role outside the one-particle sector).
%even if they do exist, Nature might not prefer them to (L.1c$^*$)

Another interesting class of alternatives to (L.1c$^*$)
are postulates involving a quadratic, rather than linear, deformation
of the dispersion relation, 
such as $E^2 = c^2 p^2 - {\tilde L}_p^2 p^2 E^2$
or $E^2 = c^2 p^2 - {\tilde L}_p^2 E^4/c^2$.
It appears that there should not be severe obstacles to finding
at least one consistent postulate of this type. It is perhaps 
worth emphasizing that in general it will not be sufficient
to analyze boosts along the direction of motion of the observed particle. 
Any claim of
logical consistency of the postulates should analyze all boosts and 
rotations and combinations of them. In the present study I could spare 
myself this most detailed consistency check, because the postulates
were evidently rotationally invariant and by imposing that boosts
along the direction of motion be consistent with the Relativity Principle
I was led to a boost generator which I recognized as  
the leading-order-in-${\tilde L}_p$
version of the generators of a relevant quantum algebra,
which reassured me that no additional consistency checks were
necessary in the one-particle sector.

The case of quadratic deformations of the dispersion relations
is also relevant for the issue of testable predictions.
As mentioned, the illustrative postulate (L.1c$^*$) provides an 
example in which the role of the Planck length in Relativity
is not merely an academic issue, since it does have observable
consequences. The fact that there are effects that can be tested in 
the near future is however not to be expected of all possible 
postulates involving the Planck length.
Replacing (L.1c$^*$), and its linearly-deformed dispersion relation,
with a postulate involving a quadratic deformation of the dispersion
relation one will naturally end up predicting much smaller effects,
probably too weak for testing in the near future.

Of course, it appears also possible to replace (L.1c$^*$)
with a postulate that predicts stronger effects. For example,
it would be interesting to find consistent postulates based on
a more complicated dispersion relation (involving a more
complicated role for the particle mass, through the function $F$
introduced in Section~3) with the property
that around $10^{20}eV$ the correction to the proton dispersion 
relation is of order ${\tilde L}_p c^{-3} E^4/m_{proton}$
or ${\tilde L}_p^2 c^{-5} E^6/m_{proton}^2$; in fact, this would provide a 
solution to the observed violations of the GZK threshold limit 
for cosmic rays, within Relativity with observer-independent $c$ 
and $L_p$.

\subsection{Renormalization}

The fact that it is possible to have observer-independent 
velocity and length scales without violating the Relativity Principle
appears to provide also a tool for developments
in certain active areas of theoretical physics.
One example is the one of renormalization theory.

Combining Quantum Mechanics and Special Relativity (the relativistic 
theory with a single observer-independent scale $c$), one is 
led to the familiar field theories.
These theories are affected by apparent divergences, which are
handled appropriately through renormalization.
An alternative way to handle the divergences is the introduction
of a ``physical cut-off": a scale that separates physical degrees of freedom 
from  other (in a sense, spurious) degrees of freedom that cause the 
divergences. However, the ``physical cut-off" alternative is not to be 
preferred for many reasons, including the fact that the types of cut-offs
that would lead to regularization are not invariant in the special-relativistic
sense. For example, one cannot cut-off degrees of freedom corresponding
to momenta higher that a certain $\Lambda$, because different observers
would then disagree on which degrees of freedom are to be considered
as physical.

It is legitimate to wonder whether replacing Special Relativity
by a relativistic theory with observer-independent
scales of both velocity and length (not necessarily the specific
illustrative example (L.1c$^*$)) one could be led to a significantly
different situation for what concerns the interplay with Quantum Mechanics.
For example, it appears plausible that the presence of
the observer-independent
length scale might ``cure" the divergences authomatically
or else, if the divergences are still present, 
allow to eliminate the divergences 
by introducing some type of ``physical cut-off" that would select
the physical degrees of freedom in a
way on which all inertial observers would agree.

\subsection{Quantum space-times}

Another example of theory subject in which a role could be played by
my observation that observer-independent velocity and length scales 
can be introduced without violating the Relativity Principle
is the one of studies of quantum space-times~\cite{gacqm100}, 
descriptions of space-time
that in one or another way involve some ingredients of Quantum Mechanics
(such as discrete variables and/or uncertainty relations).
In the study of this subject it is frequently assumed that new-physics
effects, due to quantum properties of space-time, should be strong
for particles of wavelength of the order of the Planck length
while they should be weak for particles of larger wavelengths.
If the Relativity postulates do not involve an observer-independent
length scale this idea appears to be problematic (different observers
would disagree on the relevance of quantum properties of space-time
for a given process, since the particles involved in the process would
have different wavelengths for different observers).
The introduction of an observer-independent length scale
might be useful for these issues.

Similarly, it is also common to conjecture that space-time might be
discrete with characteristic discreteness scales identified with the Planck
length, but, if there is no observer-independent length scale,
the discreteness scales (at least the one in the boost direction)
should appear subPlanckian to other inertial observers (observers in 
motion with the respect to the one that describes space-time as
discrete with distreness scale given by the Planck length).
Again it appears reasonable to conjecture that a more satisfactory
description of discrete space-times could emerge in a relativistic
framework with observer-independent scales of both velocity and length.

\section*{Acknowledgements}
Before starting the writeup of
a manuscript, I had the privilege to describe
some of these results to Lee Smolin. 
The interest he expressed in these findings provided
motivation for seeking publication. Those conversations with Lee also
led to a small but significant shift in my way to motivate
deformations of boosts. I felt at the time that the idea of deforming
boosts could be motivated using two arguments which to me appeared 
equally significant:
(A1) if some conceptual arguments happen to invite us to think that
one of the space-time coordinates is special it is natural to assume that
the special coordinate is time, an assumption which is
intuitively consistent with the nature of most of our observations;
(A2) since boosts are involved in the process 
of Fitzgerald-Lorentz-Einstein
length contraction (and time dilatation), 
it might be necessary to revise the concept of boost
in order to accommodate an observer-independent Planck length.
In the conversations with Lee it emerged that for him (A2) would be a 
stronger motivation (he had been led to consider
deformations of boosts by his own 
independent studies of Planck-length physics). 
I now believe that Lee's opinion is correct, and accordingly
in reporting the results I have here chosen to discuss my motivation
for deformation of boosts in a way that places much greater emphasis
on (A2) than on (A1).
I should also thankfully acknowledge that
at a later stage, when a rough manuscript was already in existence,
I described some of these results to Jurek Kowalski-Glikman.
Jurek's interest provided further encouragement for seeking publication
and he also brought to my attention Ref.~\cite{rueggnew} which
was useful for the preparation of Section~6.

\baselineskip 12pt plus .5pt minus .5pt

\vfil

\end{document}